\documentclass[10pt]{revtex4}

\usepackage{amsmath}
\usepackage{rotating}
\usepackage{hyperref}
\hypersetup{
	colorlinks   = true, %Colours links instead of ugly boxes
	urlcolor     = blue, %Colour for external hyperlinks
	linkcolor    = blue, %Colour of internal links
	citecolor    = cyan %Colour of citations
}

\usepackage{graphicx,epstopdf}

\begin{document}

\title{Unified first law and some general prescription: a redefinition of surface gravity }

\author{Sourav Haldar\footnote {\color{blue}sourav.math.ju@gmail.com}}
\affiliation{Department of Mathematics, Jadavpur University, Kolkata, West Bengal, 700032, India.}

\author{Sudipto Bhattacharjee\footnote {\color{blue}slg00sudipto@gmail.com}}
\affiliation{Department of Mathematics, Jadavpur University, Kolkata, West Bengal, 700032, India.}

\author{Subenoy Chakraborty\footnote {\color{blue}schakraborty.math@gmail.com}}
\affiliation{Department of Mathematics, Jadavpur University, Kolkata, West Bengal, 700032, India.}

%%%%%%%%%%%%%%%%%%%%%%%%%%%%%%%%%%%%%%%%%%%%%%%%%%%%%%%%%%%%%%%%%%%%%%%%%%%%%%%%%%%%%%%%%%%%%%%%%%%%%%%%%%%%%%%%%%%%%%%%%%%%%%%%%%%%%%%%%%%%%%%%%%%%%
%%%%%%%%%%%%%%%%%%%%%%%%%%%%%%%%%%%%%%%%%%%%%%%%%%%%%%%%%%%%%%%%%%%%%%%%%%%%%%%%%%%%%%%%%%%%%%%%%%%%%%%%%%%%%%%%%%%%%%%%%%%%%%%%%%%%%%%%%%%%%%%%%%%%%
\begin{abstract}
	The paper contains an extensive study of the unified first law\,(UFL) in the Friedmann--Robertson--Walker spacetime model.
	By projecting the UFL along the Kodama vector the second Friedmann equation can be obtained. Also studying the UFL on the
	event horizon it is found that Clausius relation cannot be obtained from the UFL by projecting it along the tangent to the
	event horizon as it can be for the trapping horizon. However, it is shown in the present work that Clausius relation can
	be obtained by projecting the UFL along the Kodama vector on the horizon and the result is found to be true for any horizon.
	Finally motivated by the Unruh temperature for the Rindler observer, surface gravity is redefined and a Clausius relation is
	obtained from the UFL by projecting it along a vector analogous to the Kodama vector.\\\\
%	{\bf Keywords\,:} Clausius relation, Hawking temperature, Bekenstein entropy, Kodama vector, Trapping horizon, Unified first law.\\\\
%	PACS Numbers\,: 98.80.Cq, 98.80.-k, 95.35.+d
\end{abstract}

%%%%%%%%%%%%%%%%%%%%%%%%%%%%%%%%%%%%%%%%%%%%%%%%%%%%%%%%%%%%%%%%%%%%%%%%%%%%%%%%%%%%%%%%%%%%%%%%%%%%%%%%%%%%%%%%%%%%%%%%%%%%%%%%%%%%%%%%%%%%%%%%%%%%%
\maketitle
%%%%%%%%%%%%%%%%%%%%%%%%%%%%%%%%%%%%%%%%%%%%%%%%%%%%%%%%%%%%%%%%%%%%%%%%%%%%%%%%%%%%%%%%%%%%%%%%%%%%%%%%%%%%%%%%%%%%%%%%%%%%%%%%%%%%%%%%%%%%%%%%%%%%%

\section{Introduction}

In the 1970s Hawking \cite{Hawking1} showed that black hole (BH) is not totally black; rather it emits thermal radiation by a combined
application of quantum mechanics and general relativity at semi-classical level. Interestingly, the temperature of
the radiation (known as the Hawking temperature) and the entropy of the horizon (known as the Bekenstein entropy) have a
certain universality in the sense that surface gravity (proportional to the Hawking temperature) and the horizon area
(proportional to the Bekenstein entropy) \cite{Hawking1, Bekenstein1} are purely geometric entity characterized by the
spacetime geometry. Also this entropy and temperature are related to the BH mass through the first law of BH
thermodynamics: \(\mathrm{{d}}M= T\mathrm{{d}}S\) \cite{Bardeen1}. Moreover, this fantastic discovery gave rise to (i) a speculation for
a deep interrelationship between gravity theories and thermodynamics and (ii) a clue to the nature of quantum gravity.

However, the first possibility came true in 1995 when Jacobson \cite{Jacobson1} derived Einstein equations from the
Clausius relation:
\(\delta Q= T \mathrm{{d}}S\) for the whole local Rindler causal horizon through a spacetime point (\(\delta Q\rightarrow \) the energy
flux, \(T\rightarrow\) Unruh temperature seen by the accelerated observer just inside the horizon). Subsequently,
Padmanabhan \cite{Padmanabhan1,Padmanabhan2} was able to show the first law of thermodynamics on the horizon,
starting from the Einstein equations, for a general static spherically symmetric spacetime.

Assuming the universe to be a thermodynamical system, this nice interrelation between Einstein equations and thermodynamic
laws has been extended in the context of cosmology. For a homogeneous and isotropic FRW model the Hawking temperature at the apparent horizon: \(T_{A}= \frac{1}{2 \pi R_{A}}\)~(\(R_A\)= geometric radius of the apparent horizon) was first derived by Cai et al. \cite{Cai:2008gw}, it was found \cite{Cai1,Akbar1,Paranjape1}
that the Friedmann equations are equivalent to the first law of thermodynamics on the apparent horizon having Hawking
temperature \(T_{A}\) and Bekenstein entropy $S_A= \frac{\pi R^2 _A}{G}$. Then in higher dimensional spacetime, this equivalence was established for gravity with the
Gauss--Bonnet term \cite{Sheykhi:2007gi,Wheeler:1985nh} and for Lovelock gravity \cite{Chakraborty:2015wma,Chakraborty:2014joa}.

On the contrary, the situation is totally different for a universe bounded by the event horizon (which exists only in an accelerating phase of the expansion). Wang et al. \cite{Wang1} showed that a universe bounded by an apparent horizon
is a perfect thermodynamical system as both first and second law of thermodynamics hold for a perfect fluid with constant
equation of state and holographic dark energy models. However, according to them both thermodynamical laws failed
to be satisfied on the event horizon. Then assuming the first law, Mazumdar and Chakraborty \cite{Mazumdar1,Mazumdar2,Mazumdar3}
were able to satisfy the second law of thermodynamics on the event horizon with some realistic restrictions. In analogy
with the apparent horizon, the entropy and temperature at the event horizon were chosen as $S_E= \frac{\pi R^2 _E}{G}$
and $T_E= \frac{1}{2\pi R_E}$. Later, it was found \cite{Chakraborty1,Saha1} that the temperature taken on the event
horizon (i. e. \(T_E= \frac{1}{2\pi R_E}\)) is not correct and taking the corrected form (i. e. \(T^{(m)} _E=
\frac{R_E}{2\pi R^2 _A}\)) the thermodynamics on the event horizon has been studied. It has been shown \cite{Chakraborty2}
that for the two choices
\begin{eqnarray} \label{eq1}
\left. \begin{array}{l}
(a)~~S^{(B)} _E= \frac{\pi R^2 _E}{G} ,~~T_E= T^{(g)} _E= \alpha T^{(m)} _E= \frac{\alpha R_E}{2\pi R^2 _A},
,~~\alpha = \frac{\dot{R_A}/R_A}{\dot{R_E}/R_E},\\
(b)~~S^{(m)} _E= \beta S^{(B)} _E,~~T_E= T^{(m)} _E ,~~\beta = \frac{2}{R^2 _E} \int R^2 _{E} \frac{dR_A}{R_A},
\end{array} \right.
\end{eqnarray}
both thermodynamical laws are satisfied on the event horizon. Also for infinitesimal thermal fluctuation, there
is a logarithmic correction to the Bekenstein entropy in the second choice \cite{Chakraborty2}. However, there are no other motivations and  proof for the above two choices in Eq. \eqref{eq1}.

On the other hand, in the context of a dynamical BH, Hayward \cite{Hayward1,Hayward2,Hayward3,Hayward4} introduced the
notion of a trapping horizon and proposed a method to deal with the thermodynamics associated with a trapping horizon.
According to him, for spherically symmetric spacetimes, the Einstein equations can be rewritten in a form termed the
``unified first law". Then projecting this UFL along a trapping horizon, the first law of
thermodynamics was derived. Further, from the point of view of universal thermodynamics we consider our universe
as a non-stationary gravitational system and the FRW model may be considered as a dynamical spherically symmetric
spacetime. Moreover, in the FRW model we have only an inner trapping horizon, which coincides with the apparent horizon
\cite{Hayward1,Hayward2,Hayward3,Hayward4,Cai2,Bak1} and the Friedmann equations are equivalent to the UFL on the
apparent horizon \cite{Cai2,Akbar2}. Also the projection of the UFL along the tangent to the apparent horizon gives the
Clausius relation \cite{Cai2}.

Further, there is no preferred time coordinate in an evolving time dependent spacetime as there is no longer any
(asymptotically timelike) Killing vector field. To resolve this problem, Kodama \cite{Kodama1} came forward with
a geometrically natural divergence-free vector field which exists in any time-dependent spherically symmetric
spacetime. This vector in the literature is commonly known as the Kodama vector, and it identifies a natural timelike direction outside a dynamic BH. Also there is a conserved current associated with this vector field
\cite{Kodama1,Hayward5}.

We have organized this paper as follows. In Section II we derive the  Friedmann equations from the UFL projecting along the Kodama
vector. In Section III we obtain the Clausius relation using the Kodama vector. Section IV leads to a redefinition of surface
gravity. Section V is devoted to conclusions.\\\\

\section{Friedmann Equations from the ufl}

The line element for FRW spacetime can be written as \cite{Trodden1}
\begin{eqnarray} \label{eq2}
ds^2 &=& h_{ab}dx^{a}dx^{b}+ R^{2}d\Omega ^2 _2, \nonumber \\
&=& -dt^{2}+ \frac{a^2}{1-kr^2}dr^{2}+ R^{2}d\Omega ^2 _2,
\end{eqnarray}
where $R= ar$ is the area radius, $h_{ab}= \mbox{diag}(-1,\frac{a^2}{1-kr^2})$ is the metric on the two-space orthogonal to the
spherical symmetry. Using null coordinates $(\xi_+,\xi_-)$ the above metric can be written as
\begin{equation} \label{eq3}
ds^2= -2\,d\xi_+\,d\xi_-+ R^{2}d\Omega ^2 _2,
\end{equation}
with
$$\frac{\partial}{\partial \xi_\pm}= -\sqrt{2}\left(\frac{\partial}{\partial t}\mp \frac{\sqrt{1 - kr^2}}{a} \frac{\partial}{\partial r}\right),$$
as future pointing null vectors.

Now according to Hayward \cite{Hayward1,Hayward2,Hayward3,Hayward4}, the trapping horizon ($R_T$) is defined as
$(\partial _{t}R)_{R= R_T}= 0$, i.e.
\begin{equation} \label{eq4}
R_T= \frac{1}{\sqrt{H^2 + \frac{k}{a^2}}}= R_A,~~~~k=0, \pm 1
\end{equation}
where $R_A$ is the geometric radius of the apparent horizon. For the trapping horizon (i.e. the apparent horizon for FRW model) the surface gravity is defined as \cite{Helou1}
\begin{equation} \label{eq5}
{\kappa}_T = \frac{1}{2\sqrt{-h}}\partial _{a}(\sqrt{-h}\,h^{ab}\partial _{b}R){\Big|}_{R=R_T}.
\end{equation}\par
However, in the present work we are examining the equivalence of the UFL and the Friedmann equations over any horizon and, consequently, we are assuming the above definition of surface gravity (i.e. Eq.\,(\ref{eq5})) is true for any horizon, i.e. for any horizon  (having area radius $R$) the surface gravity is defined as
\begin{equation} \nonumber
{\kappa} = \frac{1}{2\sqrt{-h}}\partial _{a}(\sqrt{-h}\,h^{ab}\partial _{b}R),
\end{equation}
or in explicit form
\begin{equation} \label{eq6}
\kappa = -\left(\frac{R}{R_A}\right)^{2}\left(\frac{1- \dot{R}_A/{2HR_A}}{R}\right).
\end{equation}

Now the total energy inside the horizon is a purely geometric quantity, related to the structure of the spacetime
and to the Einstein equations \cite{Helou1}. According to Misner and Sharp \cite{Hayward1,Hayward2,Hayward3,Hayward4,Cai2,Bak1},
the total energy is given by
\begin{equation} \label{eq7}
E= \frac{R}{2G}(1- h^{ab}\,\partial _{a}R\,\partial _{b}R),
\end{equation}
which on simplification gives
\begin{equation} \label{eq8}
E= \frac{R^3}{2G}\left(H^2 + \frac{k}{a^2}\right)= \frac{R^3}{2GR_A^2}.
\end{equation}
According to Hayward \cite{Hayward1,Hayward2,Hayward3,Hayward4}, the UFL,
\begin{equation} \label{eq9}
dE= A\psi + WdV,
\end{equation}
is nothing but a rearrangement of the Einstein equations. In the above, $A$ and $V$ stand
for the area and volume bounded by the horizon, and the work density,
\begin{equation} \label{eq10}
W= -\frac{1}{2}T^{ab}h_{ab},
\end{equation}
is regarded as the work done by a change of the horizon, and the energy-supply term
\begin{equation} \label{eq11}
\Psi _a= T^b _{a}\partial _{b}R+ W\partial _{a}R,
\end{equation}
determines the total energy flow (i.e. $\delta Q= A\psi$) through the horizon and $\psi= \Psi_adx^a$.\\\\
We now introduce the Kodama vector for the present FRW model. It is defined as \cite{Kodama1,Abreu1}
\begin{equation} \label{eq12}
K^a= \epsilon ^{ab}\nabla _{b}R,
\end{equation}
where $\epsilon ^{ab}$ is the usual Levi-Civita tensor in the 2D radial--temporal plane (i.e. normal to the spherical
symmetry). For the present homogeneous and isotropic FRW model
\begin{equation} \label{eq13}
\epsilon _{lm}= a(t)(dt)_{l}\wedge (dr)_{m},
\end{equation}
and
\begin{equation} \label{eq14}
K^b= \left[-a\left(\frac{\partial}{\partial t}\right)^b + HR\left(\frac{\partial}{\partial r}\right)^b\right].
\end{equation}
Note that the Kodama vector is very similar to the Killing vector $\left(\frac{\partial}{\partial t}\right)^a$ in the
de Sitter space. Also Kodama vector takes the role of the timelike Killing vector (in stationary BH spacetime)
for dynamical BH and FRW spacetime. Further, it can be used as a preferred time evolution vector field in
spherically symmetric dynamical systems.

From Eqs.\,(\ref{eq10}) and (\ref{eq11}) the explicit forms of the work density and energy-supply one form for the present model are
\begin{equation} \label{eq15}
W= \frac{1}{2}(\rho - p),
\end{equation}
and
\begin{equation} \label{eq16}
\psi = \left(\frac{\rho + p}{2}\right)\left\{-HRdt+ adr\right\}.
\end{equation}
Hence we have
\begin{equation} \label{eq17}
WdV = 2\pi R^2(\rho - p)\left\{HRdt + adr\right\},
\end{equation}
and
\begin{equation} \label{eq18}
A\psi = 2\pi R^2(\rho + p)\left\{-HRdt + adr\right\}.
\end{equation}
Also, from Eq.\,(\ref{eq8}) we obtain
\begin{equation} \label{eq19}
dE = \frac{1}{2GR_A^3}\left[R^3\left(3HR_A - 2\dot{R}_A\right)dt + 3R^2 R_A a\,dr\right].
\end{equation}
We shall now show that by projecting the UFL along the Kodama vector gives the second Friedmann equation, in general.

For the above one forms using the scalar product with the Kodama vector we have
\begin{equation} \label{eq20}
\left\langle dE , K^b \right\rangle=- \frac{aR^3 H}{G}\left(\dot{H} - \frac{k}{a^2}\right).
\end{equation}
Now
\begin{equation} \label{eq21}
\left\langle A\psi , K^b \right\rangle= 4\pi R^3 Ha(\rho + p),
\end{equation}
and
$$\left\langle WdV , K^b \right\rangle= 0.$$
Thus, projecting the UFL along the Kodama vector gives
\begin{equation} \label{eq22}
\dot{H} - \frac{k}{a^2} = -4\pi G(\rho + p),
\end{equation}
which is nothing but the second Friedmann equation on any arbitrary horizon.\\

We shall now show that the first Friedmann equation can also be obtained from the UFL by projecting it along a vector orthogonal
to the Kodama vector, namely
$$U^\mu = \left(-\frac{1}{a^2R},~\frac{1}{Ha^3R^2},~0,~0 \right).$$
Clearly the vector $U^\mu$ lies on the radial--temporal plane and it has the following properties:

(i) The vector may be spacelike, timelike or null depending on $R$.

(ii) It is divergence-free in nature (i.e. $\nabla _\mu U^\mu = 0$) and there is a current associated with the vector
$U^\mu$ given by the relation $\xi ^\mu = G^{\mu \nu}U_{\nu}.$ So from the Bianchi identity (i.e. $\nabla_{\mu}G^{\mu\nu}=0$)
the vector $\xi ^\mu$ is conserved, i.e.
$$\nabla _\mu \xi ^\mu = 0.$$
The scalar product of the individual one-form terms on both sides of the UFL with $U^\mu$ gives
\begin{equation} \label{eq23}
\left\langle dE , U^\mu \right\rangle= \frac{R^2}{2G{R_A^3}a^2}\left[-\left(3HR_A-2\dot{R}_A\right) + \frac{3R_A}{HR^2}\right],
\end{equation}
\begin{equation} \label{eq24}
\left\langle A\psi , U^\mu \right\rangle = \frac{2\pi R^2}{a^2}(\rho + p)\left\{H + \frac{1}{HR^2}\right\},
\end{equation}
and
\begin{equation} \label{eq25}
\left\langle WdV , U^\mu \right\rangle = \frac{2\pi R^2}{a^2}(\rho - p)\left\{-H + \frac{1}{HR^2}\right\}.
\end{equation}
Hence projecting the UFL along $U^\mu$ and after some algebra we obtain the first Friedmann equation, i.e.
\begin{equation} \label{eq26}
H^2 + \frac{k}{a^2} = \frac{8\pi G}{3}\rho.
\end{equation}

\section{Clausius relation from the UFL}

Further, it has been shown in the literature that the first law of BH thermodynamics can be obtained by projecting the UFL
along the trapping horizon \cite{Hayward1,Hayward2,Hayward3,Hayward4,Cai2},
i.e.
\begin{equation} \label{eq27}
\left\langle A\psi , z \right\rangle= \frac{\kappa}{8\pi G}\left\langle dA , z \right\rangle,
\end{equation}
where $z$ is a vector tangential to the trapping horizon.

We shall now show that the situation is not so easy in case of an event horizon (EH). The area radius of the EH is given by
\begin{equation} \label{eq28}
R_E= a\int _{t}^{\infty} \frac{dt}{a}.
\end{equation}
(Note that the improper integral converges for accelerating phase of the universe.)

The normal vector to the null hypersurface $R- a\int _{t}^{\infty} \frac{dt}{a}= 0$ is given by $n_a= (-1,a,0,0)$, a null vector.

From the property of the null vector, $n_a$ is also tangential to the (null) event horizon hypersurface. Then one can easily
see that the Clausius relation, i.e. Eq.\,(\ref{eq27}) is not satisfied for the event horizon. Thus the result
\cite{Hayward1,Hayward2,Hayward3,Hayward4,Cai2} namely the first law of thermodynamics on the trapping horizon, obtained by projecting the UFL along the tangent to the trapping horizon, can be generalized to any other horizon by projecting along the Kodama vector.

Note that Eq.\,(\ref{eq21}) gives the rate of energy across the horizon. Thus the energy flux across the event horizon
during infinitesimal time $dt$ is
\begin{equation} \label{eq29}
dQ= 4\pi HR^3 _{E}(\rho + p)dt,
\end{equation}
or using the second Friedmann equation ($\dot{H}- \frac{k}{a^2}= -4\pi G(\rho + p)$)
\begin{equation} \label{eq30}
dQ= -\frac{HR^3 _E}{G}\left(\dot{H} - \frac{k}{a^2}\right)dt= \frac{R_E^3}{GR_A^3}\dot{R_A}dt,
\end{equation}
where $R_A= \frac{1}{\sqrt{H^2 + \frac{k}{a^2}}}$ has been used in the last equality.\\
Recently, a notion of generalized Hawking temperature \cite{Chakraborty2} (see Eq.\,(\ref{eq1}))
\begin{equation} \label{eq31}
T^{(G)} _E= \alpha \frac{R_E}{2\pi R_A^2},
\end{equation}
has been introduced on the event horizon for the validity of the thermodynamical laws. So in the present context
using the first choice in Eq.\,(\ref{eq1}) we have
\begin{equation} \label{eq32}
T^{(G)} _{E}dS_E= \frac{R_E^3}{GR_A^3}\dot{R}_Adt.
\end{equation}

Thus we obtain the Clausius relation $\delta Q= T^{(G)} _{E}dS_E$ on the event horizon, by projecting the UFL along
the Kodama vector on the horizon. It is interesting to note that the present approach to obtain the Clausius relation (i.e. the first law of thermodynamics) from the UFL is a general prescription and it holds in any horizon
even in the trapping horizon.\\\\

\section{A redefinition of surface gravity}

Finally, we redefine the surface gravity motivated by the concept of a Rindler observer. We have seen that at the local Rindler causal horizon, the
Unruh temperature is the effective temperature experienced by a uniformly accelerating detector in a vacuum field. This temperature ($T=\frac{\hbar f}{2\pi c k_B}$) is proportional to the local acceleration ($f$) and has the same form as the Hawking temperature of a black hole \cite{misner}. So in analogy with the Unruh temperature we assume that the temperature $T_{RH}$ (the redefined Hawking temperature) should be proportional to the acceleration of the model (in units $\hbar=1=c=k_B$), i.e.
\begin{equation} \label{temperaturerinler}
T_{RH}=\frac{A}{2\pi},
\end{equation}
where the acceleration ($A$) of the FRW model (equivalently, it is the acceleration of a hypothetical observer moving with the expansion of the universe) can be defined as 
\begin{equation} \label{eq33}
A=a_0R\frac{\ddot{a}}{a}=a_0R\left(\dot{H}+H^2\right).
\end{equation}
Here, $a_0$ is a dimensionless constant of proportionality (to be determined from the present context) and $R$ is introduced on dimensional grounds. Then using the Einstein equations, the redefined Hawking temperature becomes
\begin{equation} \label{eq34}
T_{RH} = -\frac{2a_0 RG}{3}(\rho + 3p).
\end{equation}
By introducing the vector (ad hoc, in analogy with the Kodama vector, having the properties given below)
\begin{equation} \label{eq35}
V^{\alpha} = \left(-\frac{3}{R^3},~\frac{H}{aR^2},~0,~0\right),
\end{equation}
we obtain
$$\left\langle dE , V^{\alpha} \right\rangle = 4\pi H(\rho + 3p),$$
and thus we have the Clausius relation
$$\left\langle dE , V^{\alpha} \right\rangle = \left\langle T_{RH}dS , V^{\alpha} \right\rangle,$$
$$i.e.~~~\delta Q = T_{RH}dS,$$
provided $a_0 = \frac{3}{2}$ and $S$ is the usual Bekenstein entropy. Therefore, we have the Clausius relation obeying the temperature--acceleration relation due to Rindler.\par
It is worth mentioning that the vector $V^{\alpha}$ (termed as a modified Kodama vector) has properties similar to the Kodama vector namely\\
(i)~\(V^{\alpha}\) lies in the radial--temporal plane, neither parallel nor orthogonal to the Kodama vector as well as to $U^{\mu}.$\\
(ii)~Depending on the choice of $R$, the vector may be spacelike, timelike or null in nature.\\
(iii)~Due to the divergence-free nature of $V^{\alpha}$ (i.e. $\nabla_{\alpha}V^{\alpha}=0$), there exists a conserved current $C^{\alpha}=G^{\alpha\beta}V_{\beta}$ associated with it, i.e. $\nabla_{\alpha}C^{\alpha}=0$.

\section{conclusion}

So we have the following conclusions from Sections II--IV\,:

\textbf{(a)} Projecting the UFL along the Kodama vector, the second Friedmann equation is always obtained.\\

\textbf{(b)} Projecting the UFL along a vector orthogonal to Kodama vector\,(having other properties same as Kodama vector),
the first Friedmann equation can be obtained.\\

\textbf{(c)} Projecting the UFL along the tangent to the horizon to obtain the first law of thermodynamics
is valid only for the trapping horizon.\\

\textbf{(d)} First law of thermodynamics on any horizon can be obtained from the UFL by projecting
it along the Kodama vector on the horizon.\\

\textbf{(e)} Clausius relation has been obtained on any horizon considering Rindler's temperature--acceleration relation.\\

Thus we have shown that the Hayward--Kodama definition of surface gravity\,(defined in Eq.\,(\ref{eq5})) is valid for 
dynamical models. However, projecting with the Kodama vector we
cannot only obtain both Friedmann equations from the UFL but also are able to obtain the Clausius relation on the event
horizon (or any horizon) with temperature as generalized Hawking temperature. Also the first Friedmann equation can be obtained
from the UFL by projecting along the orthogonal direction of the Kodama vector. Finally, in analogy with the Unruh temperature the
surface gravity is redefined as proportional to acceleration in FRW model and it is possible to obtain the Clausius relation
by projecting the UFL along $V^{\alpha}$ (i.e. the modified Kodama vector).

%%%%%%%%%%%%%%%%%%%%%%%%%%%%%%%%%%%%%%%%%%%%%%%%%%%%%%%%%%%%%%%%%%%%%%%%%%%%%%%%%%%%%%%%%%%%%%%%%%%%%%%%%%%%%%%%%%%%%%%%%%%%%%%%%%%%%%%%%%%%%%%%%%%%%

\begin{acknowledgements}
 The authors are thankful to the anonymous referee for his/her clarifying comments. The author S. H. acknowledges UGC for awarding JRF. The author S. B. is thankful to CSIR for awarding JRF. The author S. C. is
 thankful to the Inter University Centre for Astronomy and Astrophysics\,(IUCAA), Pune, India for research facilities at Library.
 Also S. C. acknowledges the UGC-DRS Programme in the Department of Mathematics, Jadavpur University.\\\\
	
\end{acknowledgements}

%%%%%%%%%%%%%%%%%%%%%%%%%%%%%%%%%%%%%%%%%%%%%%%%%%%%%%%%%%%%%%%%%%%%%%%%%%%%%%%%%%%%%%%%%%%%%%%%%%%%%%%%%%%%%%%%%%%%%%%%%%%%%%%%%%%%%%%%%%%%%%%%%%%%%

%%%%%%%%%%%%%%%%%%%%%%%%%%%%%%%%%%%%%%%%%%%%%%%%%%%%%%%%%%%%%%%%%%%%%%%%%%%%%%%%%%%%%%%%%%%%%%%%%%%%%%%%%%%%%%%%%%%%%%%%%%%%%%%%%%%%%%%%%%%%%%%%%%%%%
	
	\normalsize

\end{document}